\documentclass[]{aa}

    \usepackage{graphicx, natbib}
    \usepackage[dvips, bookmarks=false]{hyperref}

\PassOptionsToPackage{hdvips, pdfmark, rerunfilecheck}{hyperref}

\begin{document}
\newcommand{\mum}{\ensuremath{\rm \,\mu m}}
\newcommand{\Ms}{\ensuremath{\,M_{\sun}}}
\newcommand{\Zs}{\ensuremath{\rm \,Z_{\sun}}}
\newcommand{\Mspc}{\ensuremath{\,\Ms\,\rm \rm pc^{-2}}}
\newcommand{\Msyr}{\ensuremath{\,\Ms\,\rm yr^{-1}}}
\newcommand{\pyr}{\ensuremath{\rm \,yr^{-1}}}
\newcommand{\cmc}{\ensuremath{\rm \,cm^{-3}}}
\newcommand{\kms}{\ensuremath{\rm \,km\,s^{-1}}}
\newcommand{\ddt}[1]{{{\rm d}\, {#1} \over{\rm d}\,t}}

\author{Svitlana Zhukovska\inst{1,}\inst{2} and Thomas Henning\inst{1}}
\institute{Max-Planck-Institut f\"ur Astronomie, K\"onigstuhl 17, D-69117 Heidelberg, Germany
\and 
Zentrum für Astronomie der Universit\"at Heidelberg, Institut f\"ur Theoretische Astrophysik, Albert-Ueberle-Str. 2, 69120 Heidelberg, Germany
}
\title{Dust input from AGB stars in the Large Magellanic Cloud}
\titlerunning{Dust input from AGB stars in the LMC}
\abstract
{}
{
The dust-forming population of AGB stars and their input to the interstellar dust budget of the Large Magellanic Cloud (LMC) are studied with evolutionary dust models with the main goals (1) to investigate how the amount and composition of dust from AGB stars vary over galactic history; (2) to characterise the mass and metallicity distribution of the present population of AGB stars; (3) to quantify the contribution of AGB stars of different mass and metallicity to the present stardust population in the interstellar medium (ISM).}
{
We use models of the stardust lifecycle in the ISM developed and tested for the Solar neighbourhood. The first global spatially resolved reconstruction  of the star formation history of the LMC from the Magellanic Clouds Photometric Survey is employed to calculate the stellar populations in the LMC.}
{
The dust input from AGB stars is dominated by carbon grains from stars with masses $\lesssim 4\Ms$ almost over the entire history of the LMC. The production of silicate, silicon carbide and iron dust is delayed until the ISM is enriched to about half the present metallicity in the LMC.
For the first time, theoretically calculated dust production rates of AGB stars are compared to those derived from IR observations of AGB stars for the entire galaxy. We find good agreement within scatter of various observational estimates. We show that the majority of silicate and iron grains in the present stardust population originate from a small population of intermediate-mass stars consisting of only $\lesssim 4$\% of the total number of stars, whereas in the Solar neighbourhood they originate from low-mass stars.
With models of the lifecycle of stardust grains in the ISM we confirm a large discrepancy between dust input from stars and the existing interstellar dust mass in the LMC reported in Matsuura et al. 2009.} 
{}


\date{}
\maketitle


\section{Introduction}





Tiny refractory solid particles or interstellar dust constitute an important component of the interstellar medium (ISM) both in local galaxies and objects in the young universe. Dust modifies the shape of the spectral energy distribution of galaxies by absorbing stellar radiation in the ultraviolet (UV) and re-emiting it at infrared (IR) wavelengths \cite[e.g.,][]{Draine:2003p1007, Henning:2010p7233}. Grains also influence interstellar chemistry via surface reactions, in particular the formation of H$_2$ molecules, the largest constituent of molecular gas. 
Thus, a thorough understanding of the nature and origin of dust in different environments is needed for the interpretation of  observations and modelling of physical processes in the ISM \citep{Dorschner:1995p7228}.

Low- and intermediate-mass stars during the Asymptotic Giant Branch (AGB) stage of their evolution and core-collapse Supernovae (SNe)  have long been considered the main sites of dust formation, but their contributions to the interstellar dust budget are still debated. Core-collapse SNe produce most heavy elements on a short time scale, therefore they are potentially the primary source of dust in the early universe \citep[][ and references therein]{Gall:2011hr}. However, many recent studies derive low efficiencies of dust condensation in SN ejecta from mid- and far-IR observations \citep[e.g.,][]{Ercolano:2007p2083, Rho:2008p4396, Rho:2009p4397,Temim:2012vj}, theoretical modelling of dust formation and survival in SN ejecta \citep{Bianchi:2007p2222, Kozasa:2009p1041, Cherchneff:2009p3933, Cherchneff:2010p6001} and analysis of presolar grains  with different  origins found in meteorites \citep{Zhukovska:2008bw}. Recently, far-IR and submillimeter observations with the \textit{Herschel Space Observatory} permitted the detection of larger masses of newly formed dust in SN ejecta (\citealp{Matsuura:2011ij, Gomez:2012fm}; see also \citealp{Lakicevic:2011fl} for ground-based sub-mm observations), but it is still unclear what fraction of these grains will survive destruction in the reverse shocks of SN remnants \citep[e.g.,][]{Silvia:2010p6576, Silvia:2012br}. Given the controversies with dust input from SNe, AGB stars may  be a dominant stellar source of dust production. Indeed \cite{Valiante:2009p7212} showed that even at redshift $z\sim  6.4$ the dust input from AGB stars can account for the high dust mass observed in one of the most distant quasars J114816 + 525150 \citep[but see also][]{Valiante:2011hu}. 

Models of cosmic dust and gas evolution in the Solar neighbourhood argue that the dust input from stellar sources alone can not reproduce observed interstellar dust abundances, because grains are efficiently destroyed in SN shocks \citep{Dwek:1980p490, Dwek:1998p67, Zhukovska:2008bw, Draine:2009p6616}. 
Additional grains growth in the ISM by selective mantle accretion is required by these models to reproduce the observed interstellar dust content. The mechanism of this process is poorly understood. Because of abundance constraints and the limited lifetime of molecular clouds, it depends critically on the metallicity; thus the role of stars in dust production is expected to increase at lower metallicities \citep{Zhukovska:2008p7215}.  Additionally, \cite{2011A&A...530A..44J} recently showed that stardust grains may survive in the ISM longer than previously thought \citep[][]{Jones:1994p1037, Jones:1996p6593}, which would increase the importance of AGB stars for dust production.

The Large Magellanic Cloud (LMC) is an ideal local laboratory to investigate dust input from stars at subsolar metallicities \citep[$Z_{\rm LMC} \simeq 0.3-0.5 Z_\sun$,][]{Russell:1992eu} with evolutionary dust models. Over recent years IR dust emission from stellar sources in the LMC has been extensively studied in a number of surveys \citep[e.g.,][]{Cioni:2000wp, 2001AJ....122.1844E, 2006AJ....132.2268M,  Skrutskie:2006hl, Ita:2008vl, Blum:2006ib}. 
These studies have permitted, for the first time, measurement of the dust production rate (DPR) by stars for the entire galaxy \citep{Matsuura:2009p363, Srinivasan:6p7509, Boyer:2012ck, Riebel:2012eq}. This can be used to test and constrain models of dust evolution at the present time. For our galaxy, such a global study is complicated by the large extinction in the galactic disk. 
With the integrated DPR from IR observations of evolved stars and an assumed galactic value for the grain lifetime in the ISM of 400~Myr, \cite{Matsuura:2009p363} found a significant mismatch between estimates of the accumulated dust mass from stars and the observed interstellar dust mass in the LMC (the so called ``missing dust-mass'' problem). This is an important finding and it has to be verified with a more detailed approach.

The global dust input from stars can be investigated with evolutionary dust models if the galactic star formation history and chemical enrichment history are known. The LMC is close enough to permit global and spatially resolved reconstruction of its star formation history  (SFH) with better spatial and temporal resolution than in any other galaxy. \cite{Harris:2009p7602} performed such a reconstruction for the LMC by fitting multiband photometry of 20 million stars with detailed stellar evolution isochrones. On the other hand, there have been a large number of new studies that constrain the chemical enrichment history of stellar populations of the LMC, both in the field and in clusters \citep[e.g.,][]{Piatti:2012vz, Colucci:2012hg, Carrera:2008p7783, Pompeia:2008p7743}. These studies found that the LMC has a peculiar chemical enrichment history, characterised by a fast initial enrichment of the ISM to [Fe/H]$\sim -2 \div -1.5$, followed by a gradual increase of metallicity up to the present value of [Fe/H]$\simeq -0.5$. 
%
%
Despite these advances, information on the initial mass and metallicity distribution of the present dust-forming population of evolved stars, which is essential for understanding how the dust mixture from the low- and intermediate-mass stars varies over galactic evolution, is still missing. To improve this situation, we employ recently obtained constraints on the SFH and chemical enrichment history of the LMC, together with theoretical models for dust condensation in the stellar winds of evolved stars, to construct a model of formation and survival of stardust grains in the LMC. We apply a method which was proposed and tested for the Solar neighbourhood \citep{Gail:2009p512} to the LMC with the following main purposes: (1) to investigate how the dust production rates and grain composition from AGB stars vary over galactic history; (2) to characterise the mass spectrum and metallicities of the present population of AGB stars; (3) to compare the global DPR and accumulated mass of stardust at the present time with measurements from IR observations of AGB stars and interstellar dust in the LMC; (4) to quantify the contribution of stars of different mass and metallicity to the present-day grain population taking into account grain destruction in the ISM. 

The paper is organized as follows. Sect.~\ref{sect:AGBdustyields} describes the theoretical models of dust condensation in the stellar winds of AGB stars used as input in evolutionary models, and also observational studies used for comparison of the present DPRs. Calculations of dust production rates from stellar populations of  low- and intermediate-mass stars are presented in Sect.~\ref{sect:DustRates}. In Sect.~\ref{sect:Destr} we discuss timescales of dust destruction in the LMC. Sect.~\ref{sect:AGBcontrib} describes the method used to quantify the contribution of AGB stars to the present-day dust population as a function of initial stellar mass and metallicity. In Sect.~\ref{sect:Results}, we present the results of our calculations of  production rates by AGB stars for various dust species, and individual contributions of AGB stars as a function of initial mass and metallicity. Section~\ref{sect:Discussion} discusses these results, and compares them with present-day constraints from observations. Finally, conclusions are presented in Sect.~\ref{sec:Concl}.
 
     



\section{Dust formation by low- and intermediate-mass stars}
\label{sect:AGBdustyields}
Low- and intermediate-mass stars (with initial masses $0.8 \Ms \lesssim m \lesssim 8 \Ms$\footnote{The upper mass limit is determined by maximum mass of stars that develop an electron-degenerate C-O core and end their life as White Dwarfs.}) become prominent dust factories at the final stage of evolution \citep[e.g.,][]{Molster:2010fa, Gail:2010fs, Clayton:2004dp}. Dust formation occurs in cooling stellar outflows during strong mass-loss at the thermally pulsing (TP) AGB. Although stars evolving up to the Red Giant Branch lose mass via stellar winds, the properties of stellar wind at this stage prohibit efficient dust growth \citep{Gail:2009p512}. The transition to a highly supersonic outflow occurs rather close to the star where temperatures are too high for dust formation.  In regions where temperatures become low enough for condensation, the density in the stellar outflow is already too low for substantial dust growth.  \cite{Gail:2009p512}  also considered dust condensation in stellar winds during the early AGB phase and showed that only negligible amounts of dust can be condensed at this evolutionary stage.

The chemical composition of the dust mixture from AGB stars is determined by the carbon-to-oxygen element abundance ratio in the stellar wind. In an oxygen-rich environment  (C/O<1), most carbon atoms are quickly locked up in CO molecules, leaving an excess of  oxygen atoms to form various silicates and oxides. Among them magnesium-iron-silicates are major constituents due to  abundance constraints. For a recent review on the identification of a rich circumstellar dust mineralogy from the spectroscopy of AGB stars, we refer to \cite{Molster:2010fa}. In circumstellar shells with carbon-rich chemistry  (C/O>1), amorphous carbon dust formed out of primary carbon is dominant, while silicon carbide and magnesium sulphide are minor (but widespread) dust components \citep{Zhukovska:2008gn}. Iron is expected to efficiently condense in metallic iron dust in C- and S-stars from theoretical considerations \citep{Gail:1999p2289}, but it has long remained undetected due to its featureless spectra. There is recent observational evidence that iron grains may be an important opacity source at low metallicities \citep{McDonald:2010cy, McDonald:2011eo}. Since carbon is synthesized in AGB stars, while the abundances of other elements (O, Mg, Si, Fe) available for dust formation are determined by initial stellar abundances, there is a strong dependence of the dust mixture from AGB stars on their initial metallicity \citep[e.g.,][]{Zijlstra:2006gl, Leisenring:2008p1772}. We will discuss this dependence in detail in the following.

\begin{figure}[t]
	\vspace*{-2cm}
	\hspace*{-0.7cm}
   	\includegraphics[scale=.39]{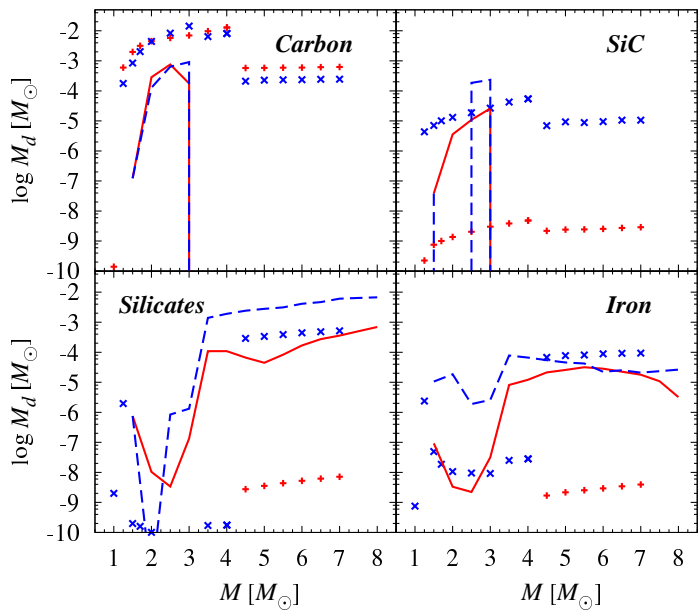}
	\caption{Dust masses for carbonaceous, silicate, SiC and iron grains condensed by AGB stars from \cite{Zhukovska:2008bw}, shown by plus symbols and crosses for initial stellar metallicities Z=0.001 and Z=0.008, respectively, and from \citet{Ventura:2012gl} (solid lines) for Z=0.001 and \cite{Ventura:2012cs} (dashed lines) for Z=0.008.}
\label{fig:dustyields}
\end{figure}

\begin{figure}[t]
	\vspace*{-2.5cm}
	\hspace*{-0.5cm}
   	\includegraphics[scale=.35]{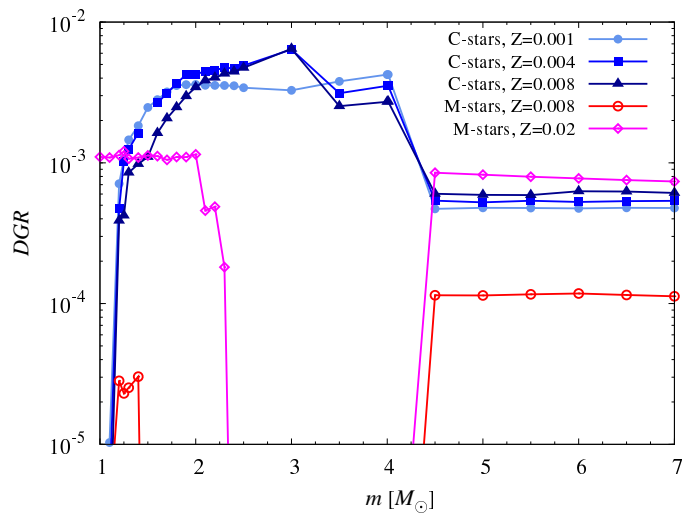}
	\caption{Dust-to-gas ratio (DGR) in the circumstellar shell at the end of M-star evolution for $Z=0.008$ and $Z=0.02$ (line with open circles and open diamonds), and at the end of C-star evolution for metallicities $Z=0.001$, 0.004, 0.008 (lines with filled circles, squares, and triangles, respectively) from calculations by \cite{Ferrarotti:2006p993} and \cite{Zhukovska:2008bw}.}
\label{fig:agbdust2gas}
\end{figure}


\subsection{Models of dust condensation in stellar winds}
\label{Sect:DustModels}

\cite{Ferrarotti:2006p993} performed extensive calculations of dust formation by low- and intermediate-mass stars during their AGB evolution for a wide range of initial stellar masses and metallicities, making these models the most suitable for use in evolutionary dust models. Their work is based on detailed theoretical models of non-equilibrium dust condensation in stellar winds, developed in a series of earlier papers \citep{Gail:1987ul, Gail:1999p2289, Anonymous:2011p5974, Ferrarotti:2002ko, Ferrarotti:2003ho, Gail:2003p2931}. These studies consider the formation of a multicomponent dust mixture in stationary, dust-driven winds with O-, C- and S-chemistries.  \cite{Ferrarotti:2006p993} combined wind models with synthetic stellar evolution models for the central star to include variations of the chemical surface composition during stellar evolution along the AGB. 

For the dust input to the ISM by low- and intermediate-mass stars, we use the mass- and metallicity-dependent dust yields from \cite{Zhukovska:2008bw}, who repeated the calculations presented in \cite{Ferrarotti:2006p993} on much finer mass and metallicity grids, and provided a detailed description of the chemical composition of  silicates.
These models include the condensation of  silicates and metallic iron for M-stars, quartz-type dust and metallic iron for S-stars, SiC and carbonaceous dust for C-stars. The composition of the silicate mixture in these models  does not vary significantly with metallicity for $Z\lesssim Z_{\rm LMC}$  and includes amorphous silicates with 40-55\% forsterite (Mg$_2$SiO$_4$), 20-35\%  enstatite (MgSiO$_3$), 15-30\%  fayalite (Fe$_2$SiO$_4$),  and $\lesssim10\%$ quartz (SiO$_2$) and ferrosilite (FeSiO$_3$) stoichiometry \citep[see Table A.1 in][for detailed composition]{Zhukovska:2008bw}. 
We combined the dust masses of these silicate types, since presently global observational studies of circumstellar dust in AGB stars in the LMC do not distinguish between them and adopt a fixed composition of silicate dust.  

Figure~\ref{fig:dustyields} shows the mass of silicate, iron, SiC and carbonaceous dust condensed in stellar outflows over the entire AGB evolution as a function of initial stellar mass for $Z=0.001$ and 0.008 from calculations by  \cite{Zhukovska:2008bw} and \cite{Ferrarotti:2006p993}. Dust-to-gas ratios in circumstellar shells at the end of stellar evolution as M-stars with initial metallicity $Z=0.008$ and as C-stars with $Z=0.001$, 0.004 and 0.008 from their calculations are shown in Fig.~\ref{fig:agbdust2gas}. The values of the dust-to-gas ratio in the outflows of O-rich stars with metallicities $Z=0.001$ and 0.004 are below $10^{-5}$ and are not shown here. Such values are probably too low, since these models may underestimate  efficiency  of dust condensation in stellar outflows from M-stars with low mass-loss rates as will be discussed further in this Section. The amounts of silicate, iron and SiC dust returned by AGB stars with $Z=0.001$ are several orders of magnitude lower than those returned by stars with $Z=0.008$, while carbon dust has comparable masses. Such strong variations in the dust mixture in these models result from (1) the increase of dust-forming element abundances with metallicity, and (2) the metallicity dependence of stellar evolution. 

The effect of reduced abundances of refractory elements at lower metallicities is most clearly seen for SiC and carbon dust, which both condense in carbon-rich environments. They have similar condensation degrees of key elements in dust for $Z=0.008$. For $Z=0.001$, the condensation degree for carbon dust remains at the same level, while for SiC it drops by about two order of magnitude. 
This can be explained with a simplified growth model, which demonstrates that, for the observed rates of mass loss in AGB stars, the grain radius attained after condensation in stellar wind is always smaller than the maximum radius corresponding to the complete condensation, and is inversely proportional to the abundance of the key element $\epsilon_{\rm el}$ \citep[see Eqs.~(13--16) in ][]{Zhukovska:2008gn}. Therefore, the volume of condensed material is proportional to $\epsilon_{\rm el}^{-3}$. This corresponds to 512 times larger SiC dust mass for $Z=0.008$, compared to that for $Z=0.001$,  if the Si abundance is scaled with metallicity and  other parameters of the stellar wind are fixed. Much higher mass-loss rates of the order of $10^{-3}\Msyr$ are required to reach the same condensation degrees in outflows of stars with $Z=0.001$. For comparison, the characteristic values of mass loss during the superwind phase at the tip of AGB are $2-3\times 10^{-5}\Msyr$ \citep{Schroder:1999ti}.
Since carbon atoms are synthesised by the star itself, carbon dust mass does not show such strong dependence on initial metallicity.

Metal-poor low-mass stars become carbon-rich after only a few thermal pulses, and thus contribute very little to oxygen-rich dust production. However, the models of \cite{Ferrarotti:2006p993} may also somewhat underestimate the amount of dust condensed in outflows from M-stars. Recently it was found that grain drift is  more important for the dust growth in outflows of AGB stars with low mass-loss rates than previously thought (Gail\&Nowotny, in preparation). Since O-rich AGB stars have generally low mass-loss rates except for brief periods related to the luminosity spikes following a thermal pulse \citep[e.g., Fig.5 in][]{Ferrarotti:2006p993}, including grain drift in the models may increase the total mass of dust output  by M-stars. For C-stars, the impact of grain drift is expected to be small, because these stars experience higher mass loss. A quantitative estimate of the effect of grain drift on dust yields from AGB stars requires new calculations of dust condensation with improved models coupled with stellar evolution  to be done in future (H.-P.~Gail, private communication).

Intermediate-mass stars experience hot bottom burning that, to some extent, counteracts carbon enrichment by the third dredge-up process. Compared to stars with $Z=0.001$, stars with $Z = 0.008$ experience the significantly higher number of thermal pulses before the C/O ratio on their surface exceeds unity. The opacity of the dust mixture condensed during the oxygen-rich evolution of these stars is sufficient to drive the stellar winds, with the exception of brief periods related to the luminosity dips following thermal pulses  \citep{Ferrarotti:2006p993}. Note that stars with $Z\gtrsim 0.02$ experience efficient hot bottom burning and remain oxygen-rich for most of their TP-AGB evolution, thus producing significantly higher quantities of O-rich dust. 

In intermediate-mass stars with $Z\gtrsim 0.001$, the mass-loss rates are well below the critical value for the complete condensation of key growth species (Si atoms) in dust resulting in very low condensation degrees. Unlike in O-rich stars with $Z = 0.008$, the amount of silicate dust condensed in the circumstellar shells of these stars is not sufficient for radiation pressure to overcome the gravitational pull of the star. Only when these stars enter the carbon-rich stage of their evolution, they condense carbon dust in sufficient amounts for the dust-driven stellar winds. Stars with $Z = 0.001$ becomes carbon-rich earlier and lose a larger fraction of envelope by C-rich winds. Therefore they return somewhat larger masses of carbon dust compared to stars with higher metallicity.

Recently, \citet{Ventura:2012gl} modelled the mass and composition of dust in stellar winds of AGB stars in the mass range $1\Ms \leq M \leq 8\Ms$ with initial metallicity of 0.001, using a  more detailed treatment of stellar evolution. Calculations for stars with metallicity of the LMC ($Z=0.008$) were performed by \cite{Ventura:2012cs}. The dust masses produced over the AGB evolution resulting from their calculations are also shown in Fig.~\ref{fig:dustyields}.
The dust condensation models in these studies are similar to those presented by \cite{Ferrarotti:2006p993}. The main difference, which has a strong impact on the chemical composition of dust from AGB stars, is in the stellar evolution modelling, in particular the adopted prescription of convective transport in \citet{Ventura:2012cs,Ventura:2012gl}. This leads to a strong hot bottom burning process experienced at the bottom of the convective envelope, and prevents intermediate-mass stars from entering the carbon-rich stage and contributing to carbon dust production.  Consequently, dust input integrated over the stellar mass range of AGB stars with metallicities of the LMC is dominated by oxygen-rich dust. In contrast, the dust mixture from AGB stars calculated by \cite{Ferrarotti:2006p993} remains carbon-rich dust at this metallicity, in agreement with the findings of IR surveys of dust-forming population of evolved stars in the LMC, which are discussed in the following Section. 

\begin{table*}
\caption{Dust production rates of C-rich, O-rich and extreme AGB stars in units $10^{-6}\Msyr$ and number of these stars}
\label{tab:LMC_DPR}
\begin{tabular}{l l l l l l l}
\hline\hline
\noalign{\smallskip}
DPR, C-rich  	&   DPR, O-rich     & DPR, extreme   &  $N_{\rm C}$&  $N_{\rm O}$&  $N_{\rm X}$ & Reference  \\
AGB stars          &   AGB stars            & AGB stars  & & & &\\
\hline\noalign{\smallskip}
43 (up to 100)   &   $\gg 0.4$ &     &  1~779    &  & &\cite{Matsuura:2009p363}    \\
2.4                   &    1.4           & 23.6   &  5~800 & 8~200  & 1~400 & \cite{Srinivasan:6p7509}   \\
0.87                 &   0.95\tablefootmark{a}       &     8.62    & 5~190  & 15~243\tablefootmark{a}   &  886 & \cite{Boyer:2012ck} \\
13.0-14.2       &   5.2-5.7  & 15.1 - 16.3\tablefootmark{b}   & 6709  & 19~566   & 1~340 & \cite{Riebel:2012eq} \\
  57                   &      1.3  &   & 9~400 & 17~500 &  & This paper, dust yields \\
                         &             &   &   &  &  & from \cite{Zhukovska:2008bw}    \\
\hline
\end{tabular}
\tablefoot{\tablefoottext{a}{For O-rich AGB stars from \cite{Boyer:2012ck} we added the total number of anomalous O-rich AGB stars (6~372) and DPR of $0.32\times 10^{-6}\Msyr$ from these stars to those of the regular O-rich AGB stars.}
\tablefoottext{b}{The category "extreme stars" in \cite{Riebel:2012eq} is a subset of O-rich and C-rich AGB stars.} 
}
\end{table*}

\subsection{Observational studies of dust input from AGB stars in the LMC}
The global dust production rate for a galaxy can be determined observationally if it is measured individually for the entire stellar population of dust-forming stars. Recently, this became possible with the \textit{Spitzer} Legacy Program ``Surveying the Agents of Galaxy Evolution in the LMC'' (SAGE-LMC) which catalogued over 8 million infrared sources \citep{2006AJ....132.2268M}. 
There have been a number of recent studies which estimated the global dust production rate by evolved stars based on the SAGE-LMC program summarised in Table~\ref{tab:LMC_DPR}. \cite{Srinivasan:6p7509}, \cite{Boyer:2012ck} and \cite{Riebel:2012eq} used the same data set based on SAGE-LMC point sources complemented by optical photometry from the Magellanic Clouds Photometric Survey \citep[MCPS; ][]{Harris:2009p7602}, and near-infrared photometry from the Two Micron All Sky Survey \citep{Skrutskie:2006hl} or the InfraRed Survey Facility \citep{Kato:2007vx}. The differences in their derived DPR arise from various source classifications and approaches outlined below.

\cite{Srinivasan:6p7509} used empirical relations describing excess dust emission in the mid-IR to estimate the dust mass-loss rate from AGB stars. The IR excess from circumstellar shells is calculated by comparing the total flux from the source observed in each band to the flux expected from the central star. 
They found that the dust input from evolved stars in the LMC is dominated by highly evolved AGB stars experiencing heavy mass-loss and efficient dust condensation, "extreme AGB stars". \cite{Srinivasan:6p7509} found that these stars eject dust at a rate of $2.36\times10^{-5}\Msyr$, while low and moderately obscured C-rich and O-rich AGB stars inject dust about ten times slower. Because of their heavy obscuration, it is not possible to  separate these stars into those with O-rich and C-rich chemistries based on photometry alone, but a spectroscopic analysis for their subsample indicates that most of them are C-rich \citep[see][and references therein]{Srinivasan:6p7509}. 
Recently, \cite{Boyer:2012ck} recomputed  the global DPRs in the LMC using the same relations between dust mass-loss rate and $8\mum$ excess with a different calibration. They confirmed that extreme AGB stars dominate dust production, although their DPR is a factor of three smaller than that derived by \cite{Srinivasan:6p7509}.

An alternative, robust method to estimate the DPR of an AGB star is to fit its spectral energy distribution with a pre-computed grid of radiative transfer models of evolved stars. Such Grid of Red Supergiant and Asymptotic Giant Branch ModelS (GRAMS) was designed for carbon-rich stars by \cite{Srinivasan:2011jb}, and for oxygen-rich stars by \cite{Sargent:2011p7506}. The GRAMS grid was computed for  circumstellar shells with a fixed dust mixture, which consists of 90\% amorphous carbon and 10\% SiC for C-rich chemistry and oxygen-deficient silicates from \cite{Ossenkopf:1992p7790} for O-rich chemistry. \cite{Riebel:2012eq} applied  the GRAMS model grid to fit multi-band photometry of $\sim30\,000$ AGB and Red Supergiant stars in the LMC and determine their individual DPRs and bolometric luminosities. Unlike previous studies which classify sources as O-rich or C- rich AGB candidates based on their location in the color-magnitude diagram, \cite{Riebel:2012eq} discriminate between O- and C-rich stars based on the chemistry type of the best fit GRAMS model. Their results confirmed that most of extreme AGB stars (97\%) are indeed C-rich, and that they dominate the dust input from evolved stars (75\% of total DPR), in agreement with other studies \citep{Matsuura:2009p363, Srinivasan:6p7509, Boyer:2012ck}.

\cite{Matsuura:2009p363} estimated the DPR using an empirical relation between the infrared colors and mass-loss rates of AGB stars and an alternative source classification compared to that used in studies discussed above. The value of the integrated dust mass-loss rate of $>4\times10^{-5}\Msyr$ derived by \cite{Matsuura:2009p363} for carbon-rich stars  is somewhat higher compared to the other works, and arises from the different optical constants used to estimate the individual stellar DPRs and calibrate the IR colors -- DPR relation. For the DPR of oxygen-rich stars, they provide a lower limit of $4\times10^{-7}\Msyr$.

\section{Calculation of stardust production rates}
\label{sect:DustRates}

In this section we describe how we calculate dust injection rates from different stellar populations using the dust yields discussed in the previous section. If the galactic star formation history $\psi$ is known, the total production rate of stardust of type $j$ is determined by integration over all masses of low- and intermediate-mass stars ending their life at time $t$
\begin{equation}
	\dot{M}^j_{\rm d}(t) =  
	 m_{\rm av}^{-1} \int_{m_l}^{m_{\rm WD}}{\phi(m) \psi  \left(t-\tau(m,Z)\right) Y_{j, \rm d}(m,Z)dm},
	\label{eq:DustInjRate} 
\end{equation}
where $\phi(m)$ is the initial stellar mass function (IMF), $m_{\rm av} = \int m \phi(m) dm$ is the average stellar mass, $\psi$ is the star formation rate in units \Msyr, $\tau(m,Z)$ is the stellar lifetime, and $Y_{j,\rm d}(m,Z)$ is the mass of dust returned to the ISM by a star with mass $m$ and metallicity $Z$ (dust yield). The lower integration limit is determined by a stellar mass with the lifetime which satisfies $\tau(m_l)=t$ and the upper mass-limit $m_{\rm WD}=8\Ms$ is determined by stars becoming AGB-stars.

Similarly, the total (gas+dust) injection rate of element $i$ is 
\begin{equation}
	\dot{M}^i_{\rm tot}(t) =  
	 m_{\rm av}^{-1} \int_{m_l}^{m_{\rm WD}}{\phi(m) \psi  \left(t-\tau(m,Z)\right) Y_{i}(m,Z)dm},
	\label{eq:GastInjRate} 
\end{equation}
where $Y_{i}(m,Z)$ is the stellar yield of element $i$, which represents the mass of this element returned to the ISM by a star of mass $m$ and metallicity $Z$.

\begin{figure}[t]
	\vspace*{-2.5cm}
   	\includegraphics[scale=.35]{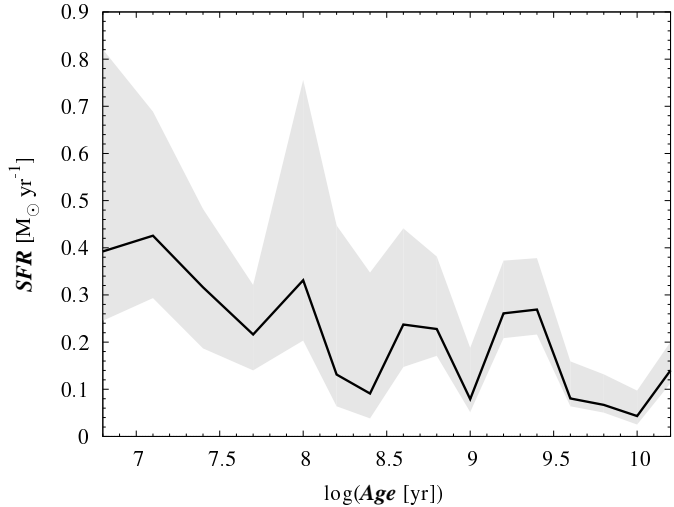}
	\caption{Variations of the total star formation history of the LMC with the age of stellar populations reconstructed by \cite{Harris:2009p7602}. The best-fit SFR as a function of age is shown with a thick black line, and the uncertainty on the fit is shown as a gray shaded area.}
\label{fig:SFR}
\end{figure}

The most important factors determining the global dust production rate by stars (DPR) are the IMF and star formation history (SFH). We employ the SFH in the LMC which was reconstructed using the multiband photometry of 20 millions of its stars from the MCPS \citep{Harris:2009p7602} (Fig.~\ref{fig:SFR}). This is the first ever spatially resolved reconstruction of the star formation history for the entire galaxy. Since we are interested in the global DPR, we sum over SFHs of individual regions for each age bin. A conspicuous feature of the SFH in the LMC is the ``Age Gap'', a quiescent epoch in star formation between 10 and 4-5~Gyr ago with a star formation rate (SFR) of only 0.02\Msyr (in Fig.~\ref{fig:SFR} it can be seen between log(Age)=9.6 and 10). After the Age Gap star formation resumed at an average SFR of 0.2\Msyr.

The SFH was derived with the StarFISH algorithm, which finds the best fit to an observed color-magnitude diagram (CMD) as a linear combination of synthetic CMDs, each of which represents the predicted distribution in color and brightness for a stellar population of a particular age and metallicity. StarFISH optimizes the age and metallicity resolution provided by the base isochrones being appropriate for the observed data. \cite{Harris:2009p7602} found that four metallicities (Z=0.001, 0.0025, 0.004 and 0.008) are optimal for MCPS data. In reality, the metallicity in the ISM has been increasing slowly since the initial burst of star formation \citep{Colucci:2012hg}. Therefore, for the calculation of stardust production rates we adopt the metallicity evolution of the gas in the LMC from more accurate spectroscopic determinations of stellar ages and metallicities of field stars from the LMC disk from \cite{Carrera:2008p7783}. 
A self-consistent model of the chemical evolution of gas and dust in the LMC will be presented elsewhere (Zhukovska et al, in preparation).
Unlike in the Milky Way disk, both stellar clusters and the field population in the LMC disk show almost no metallicity gradient, with the exception of the outermost regions containing the oldest stars \citep{Pagel:1978vb, Carrera:2008p7783}. Therefore, the evolution of the LMC disk can be treated in the one-zone approximation assuming an instantaneous element  mixing within the disk.

Following \cite{Harris:2009p7602}, we adopt the Salpeter IMF,
\begin{equation}
 	\phi(m)=Cm^{-2.35}.
\end{equation}
Normalisation within the mass range $m_l=0.1\Ms$ and $m_u=100\Ms$ yields $C\approx0.06$. 

\paragraph*{Input parameters.}
For calculating $\dot{M}^i_{\rm d}(t)$ we use $m$- and $Z$-dependent dust yields, $Y_{j,\rm d}(m,Z)$, described in Sect.~\ref{sect:AGBdustyields}. We implement the stellar lifetimes as a function of initial stellar mass and metallicity by using the analytical formula from \cite{Raiteri:1996uu}, derived by fitting the results of stellar evolution calculations by  the Padova group. For calculations of the total (gas+dust) return from low- and intermediate-mass stars, we adopt the nucleosynthesis prescriptions for H, $^4$He, $^{12}$C, $^{13}$C, $^{14}$N, and $^{16}$O from work of \cite{1997A&AS..123..305V}, calculated for the range 0.8 – 8~\Ms{} of initial masses and for metallicities 0.001, 0.004, 0.008, 0.02, and 0.04. For $^{23}$Na, $^{24}$Mg, $^{25}$Mg, $^{26}$Mg, $^{26}$Al, and $^{27}$Al, we adopt the yields of \cite{Karakas:2003el} calculated for the mass-range 1.0 – 6.5~\Ms{} and metallicity range Z = 0.004, 0.008, 0.02.
 The age of the LMC is assumed to be 13~Gyr \citep{Bekki:2012ge}. Since the oldest age bin in \cite{Harris:2009p7602} extends to 15.84~Gyr because of the large uncertainties in the ages of the oldest stars, we restrict it to 13~Gyr, and rescale the star formation rate in this bin to preserve the total stellar mass formed. The results of our calculations of dust production rates are presented in Sect.~\ref{sect:Results-DPRs}


\section{Dust destruction in the ISM}
\label{sect:Destr}
In this section we examine the time scales of dust destruction for interstellar conditions in the LMC. Over cosmic time stars continuously enrich the ISM with dust grains, but only a fraction of these grains survives until the present time because of efficient dust destruction by sputtering in interstellar SN blastwaves \citep[][]{Jones:1994p1037, Jones:1996p6593}\footnote{Recently \cite{2011A&A...530A..44J} suggested that clumpiness of the ISM can reduce the exposure of grains to the SN processing and prolong their lifetimes in the ISM. Presently we neglect this possibility.}. 
The structure of the interstellar gas in the LMC is dominated by giant and supergiant shells created by multiple SNe \citep{Kim:1999hq}. Extinction properties of dust measured inside and outside of supergiant shells in the LMC show significant differences attributed to processing of dust in the ISM \citep{Misselt:4p7511}. Moreover, a high dust destruction efficiency by core-collapse SNe is required to explain the observed dust deficit near supernova remnants (SNRs) \citep{Williams:2006p1023}. Significant grain processing in the ISM is also supported by variations of the interstellar element depletions along different sightlines in the LMC \citep{Welty:5p7696, Cox:2006gc}. 

Single SNe or first SNe in a stellar cluster are most efficient for the interstellar dust destruction, because subsequent SNe in the cluster explode in a hot bubble, where dust has been already destroyed by previous SN explosions. Their blast waves are slowed down by a thick shell of swept-up material before it runs over unprocessed interstellar dust \citep{McKee:1989p1030}. 
The destruction time scale of dust species of type $j$ by SNe in the ISM can be characterised in terms of $m_{\rm cl,j}$, the mass of gas that is completely cleared of dust by a single SN explosion \citep{McKee:1989p1030}
\begin{equation}
	\tau_{{\rm d},j} = \frac{M_{\rm ISM}}{m_{{\rm cl},j} f_{\rm SN}R_{\rm SN}}\,,
	\label{eq:TauDestrGlob}
\end{equation}
where $M_{\rm ISM}$ is the total gas mass, $f_{\rm SN}$ is the fraction of SNe  that explode as single SNe within the galactic disk and destroy dust, $R_{\rm SN}$ is the total rate of SNe of type II. The quantity $m_{{\rm cl},j}$ is determined by the properties of the dust material and the structure of the blast wave
\begin{equation}
 m_{{\rm cl},j}(n_0) = \int_{v_0}^{v_f} \epsilon_j(v_s,n_0) \left| \frac{dM_s(v_s,n_0)}{dv_s} \right| d v_s\,,
	\label{eq:MassCleared}
\end{equation} 
where $v_0$ and $v_f$ are the initial and final velocities of the SNR expanding into an ambient medium of density $n_0$, respectively, $\left| \frac{dM_s(v_s,n_0)}{dv_s} \right| dv_s$ is the mass of gas swept up by a shock with velocity in the range of $[v_s,v_s+dv_s]$, and $\epsilon_j$ is the efficiency of dust destruction  in a SN shock with expansion velocity $v_s$.  One can calculate $\left| \frac{dM_s(v_s,n_0)}{dv_s} \right|$ from analytical solutions for the SNR evolution expanding in a homogeneous medium. We adopt the expression given by \cite{Dwek:2007p496} describing the adiabatic expansion and pressure-driven snow plow stages of the SNR evolution. 

For the destruction efficiency $\epsilon_j$ of carbonaceous and silicate dust, we use analytical fits to the data from extensive theoretical calculations of grain destruction in SN shocks from \cite{Jones:1994p1037}. They assume that interstellar carbon grains have graphitic properties. Recently, \cite{SerraDiazCano:2008p588} suggested that the less resilient hydrogenated amorphous carbon is more probable candidate for carbonaceous material in the ISM than graphite. However, graphite suits better for thermally condensed grains from AGB stars, which is also evident from numerous graphitic grains of AGB origin identified by their isotopic anomalies in meteorites \citep[][and references therein]{CROAT:2005kp, Zinner:2006fo}. Unfortunately, \cite{Jones:1994p1037} did not present the corresponding data for iron and silicon carbide dust, but their results for the sputtered dust mass fraction for different shock velocities for iron and for silicon carbide dust are somewhat higher but, in principle, similar to carbon dust. Therefore, we take the same $\epsilon_j$ for iron and for silicon carbide as for carbon grains. 

 In the following we estimate the quantities entering Eq.~(\ref{eq:TauDestrGlob}) and (\ref{eq:MassCleared})  for the interstellar environment of the LMC and calculate destruction time scales for the main stardust species.

\paragraph*{Effect of ambient density.}
The LMC belongs to the class of dwarf irregular galaxies. It has a thicker HI disk and larger H~I holes than those in the Milky Way which is ascribed to the lower ambient density \citep{Brinks:2002p1750}. Using observations of bow shocks around run-away stars, \cite{Gvaramadze:2010wk} get a constraint on the number density of the ambient ISM: $n_0 \lesssim 0.1\cmc$. With the formula for the gas mass swept-up by a SN shock $M_s(v_s,n_0)$ from \cite{Dwek:2007p496} and $n_0 = 0.1\cmc$,  Eq.~(\ref{eq:MassCleared}) yields $m_{\rm cl, g} \simeq 1600 \Ms$ and $m_{\rm cl, s} \simeq 1970\Ms$, for graphite and silicate grains, respectively. If we include the metallicity dependence of the velocity at the transition of SNR from adiabatic expansion to the pressure-driven snowplough stage \citep{1988ApJ...334..252C},  we derive for $Z=0.5 \Zs$ in the LMC somewhat smaller masses, $m_{\rm cl, g} \simeq 1300  \Ms$ and $m_{\rm cl, s} \simeq  1600 \Ms$. 

\paragraph*{Fraction of SN destroying dust.}
To estimate the fraction of SNe~II that destroy dust, $f_{\rm SN}$, we rely on an observational study of OB stars in the LMC that found that 20-25\% of the massive stars in the LMC belong to the field stellar population \citep{Oey:2011ts}. 
Analysing reddening of OB stars in the field and in associations \cite{Harris:1997ka} concluded that the scale height of interstellar dust is twice that of OB stars. Therefore we assume that all massive stars in the field destroy dust. Assuming that each massive star in an association destroys 10\% of dust mass destroyed by a single SN II, we derive $f_{\rm SN} \simeq 0.3$, which is very close to 0.35 estimated for our galaxy \citep{McKee:1989p1030}. 

\paragraph*{Estimates of dust destruction time scales in the LMC.} 
The total gas mass needed for calculations of the destruction time scales can be estimated from observations of H I and H$_2$ gas. We adopt a H~I mass of $4.8 \pm 0.2 \times 10^8\Ms$ from \cite{StaveleySmith:2003bl} and H$_2$ mass of  $1.0\pm0.3 \times 10^8\Ms$ from \cite{Israel:1997tm}, which yields $M(\rm HI+H_2) = 5.8 \pm 0.5 \times 10^8 \Ms$ and, correcting for He mass, $M_{\rm ISM} \simeq 7.9 \pm 0.7 \times 10^8\Ms$.
The current rate of SN II, $R_{\rm SNII}$, is determined by the adopted IMF and recent star formation rate. For the best-fit SFR from \cite{Harris:2009p7602}, the value of $R_{\rm SNII}$ is $2 \times10^{-3}\pyr$, with large variations for the lower and upper limits for the SFR, of 1.5 and $4.0 \times10^{-3}\pyr$, respectively. Large uncertainties in $R_{\rm SNII}$ are also present in estimates from SNR surveys  ($8-12\times 10^{-3}\pyr$ in \cite{Filipovi:1998ix} and $2.5 - 4.6 \times 10^{-3}\pyr$ in \cite{Maoz:2010p7732} for total core-collapse+SN Ia). 
 For the value of $2 \times10^{-3}\pyr$  for $R_{\rm SNII}$, we estimate a time scale of dust destruction from Eq.~(\ref{eq:TauDestrGlob}) for  silicate grains of $\tau^{\rm LMC}_{\rm d,s} = 0.8$~Gyr and for carbonaceous grains of $\tau^{\rm LMC}_{\rm d,c} = 1.1$~Gyr. Note that the values of the dust lifetimes calculated for the upper limit of the SFR are very close to those in the Milky Way, $\tau^{\rm LMC}_{\rm d,s} = 0.4$~Gyr and $\tau^{\rm LMC}_{\rm d,c} = 0.55$~Gyr, respectively. For iron and SiC grains, we adopt the same time scales as for carbonaceous grains. 
 
\paragraph*{Destruction by star formation process.} 
Even in the absence of SN dust destruction, accumulation of stardust grains over cosmic time is not possible because of the matter cycle between gas and stars. The current gas reservoir will be consumed by the star formation process with a time scale of
\begin{equation}
	\tau_{\rm SF} = \frac{M_{\rm ISM}}{\psi}\,.
	\label{eq:tau_sf}
\end{equation}
The best-fit value for the current SFR from \cite{Harris:2009p7602} of $0.4\Msyr$ is somewhat high compared to estimates from the integrated global ultraviolet and IR flux from the LMC \citep[$\sim 0.05-0.25\Msyr$;][and references therein]{Whitney:2008p1780}. For estimating the grain lifetime against destruction in star formation, we prefer to use more accurate estimates of the current SFR from the integrated global fluxes and adopt the upper limit value of $\psi = 0.25$\Msyr. 
For this SFR and the current gas mass of $M_{\rm ISM}=7.8\times 10^8\Ms$, the time scale of gas consumption by star formation, $\tau_{\rm SF}$ is 3.1~Gyr. This value is longer than grain lifetimes against destruction in SN blastwaves, but still much shorter than the age of the LMC.

\section{Contribution of AGB stars to the grain population}
\label{sect:AGBcontrib}

In this section we give a short description of the method which we use to characterise the population of dust-forming  stars of different mass and metallicity contributing to the present-day stardust population in the ISM.  
The initial metallicity of a star ending its life at the present time depends on the stellar lifetime: massive stars have short lifetimes and therefore their metallicities are close to the present-day metallicity of the ISM, while ages of low- and intermediate-mass stars vary from $\gtrsim 40$~Myr up to the age of the LMC, and their metallicities range from the present day value to very metal poor for the oldest stars. The initial metallicity of a star is given by metallicity of the ISM at instant of stellar birth:
\begin{equation}
\label{eq:Z_star}
    Z = Z_{\rm ISM} ( t_{\rm d} - \tau(m,Z))
\end{equation}

\cite{Gail:2009p512} proposed a method to characterise the mass distribution of AGB stars with different masses and metallicities along with their contribution to the stardust population. Here we provide a brief description of the most important quantities and refer the reader to \cite{Gail:2009p512} for detailed derivations. 
 
The average number of AGB stars ending their life during the period $t_d ... t_d+\Delta t_{\rm AGB}$ per unit stellar mass can be characterised by a mass distribution function
\begin{equation}
\label{eq:FreqAGB}
	f_{\rm AGB}(m,t_{\rm d}) = \psi(t_{\rm d} - \tau(m,Z)) \Delta t_{\rm AGB}(m,Z) \phi(m) m_{\rm av}^{-1} ,
\end{equation} 
where $\Delta t_{\rm AGB}$ is the duration of the dust-forming phase on the AGB. Efficient dust condensation occurs during the thermal pulsation phase and lasts typically 500,000 yr to a few $10^6$~yr. Equation~(\ref{eq:FreqAGB}) gives the mass spectrum of carbon-rich stars, if the duration of carbon-rich evolution is taken for $\Delta t_{\rm AGB}$. The values of $\Delta t_{\rm AGB}$ used in our calculations are discussed at the end of this section. Note that \cite{Gail:2009p512} considered the stellar birthrate $B$ per unit area and correspondingly $f_{\rm AGB}(m,t)$ had units $\Ms^{-1} pc^{-2}$. In this paper we model LMC as one zone, and $f_{\rm AGB}(m,t)$ is the integrated value over galactic surface area.

The total number of AGB stars at time $t_{\rm d}$ is simply
\begin{equation}
\label{eq:NAGB}
	N_{\rm AGB}^{\rm tot}(t_{\rm d}) = \int_{m_l}^{m_{\rm WD}}  f_{\rm AGB}(m,t_{\rm d}) dm.
\end{equation} 

The total mass of dust of type $j$ produced by AGB stars from the mass range $m...m+\Delta m$ ending their life at instant $t_d$ is then
\begin{equation}
\label{eq:DustMassAGB}
		f_{\rm AGB}(m,t_{\rm d}) Y_{j,\rm d}(m,Z) \Delta m \,.
\end{equation} 
To account for the destruction of grains during the period they spend in the ISM, the dust mass from Eq.~(\ref{eq:DustMassAGB}) has to be multiplied by the probability that a grain injected at some instant $t_{\rm d}$ into the ISM has survived until a later time $t$: 
\begin{equation}
	P_j(t_{\rm d}, t) = \exp \left(-\int_{t_{\rm d}}^t {dt'\over \tau_{j, \rm dest} (t')} \right).
\end{equation}
If $\tau_{j, \rm dest}$ is relatively short in comparison to the time scale of galactic evolution,  $P_j(t_{\rm d}, t) \simeq \exp{(-(t-t_d)/\tau_{j,\rm dest})}$

\begin{figure}[tb]
	\vspace*{-3cm}
	\includegraphics[scale=.37]{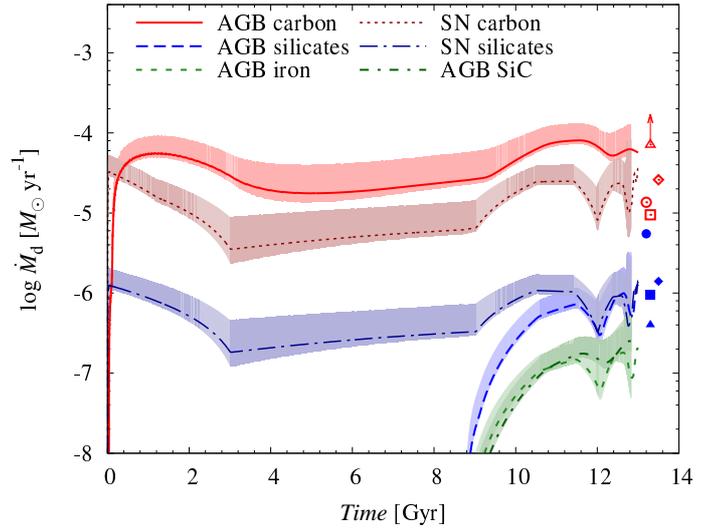}
    \caption{Variations  of dust production rates by AGB stars for silicate, carbonaceous, SiC and iron dust over the LMC evolution calculated with dust yields from \cite{Zhukovska:2008bw}. The DPR for silicates and carbonaceous grains from SN~II are shown for illustration purpose. The production rate for metallic iron by SN~II is added to that of silicates. Uncertainties in DPR for each species due to the uncertainties in the SFH fit are shown as shaded regions. Open and closed symbols show the current DPRs for carbon- and oxygen-rich dust, respectively, derived from IR observations by \cite{Matsuura:2009p363} (triangles, with an arrow marking an upper limit), \cite{Boyer:2012ck} (squares), \cite{Riebel:2012eq} (circles) and \cite{Srinivasan:6p7509} (diamonds).}
	\label{fig:DustRates}
\end{figure}

\begin{figure}[tb]
	\vspace*{-3cm}
	\includegraphics[scale=.37]{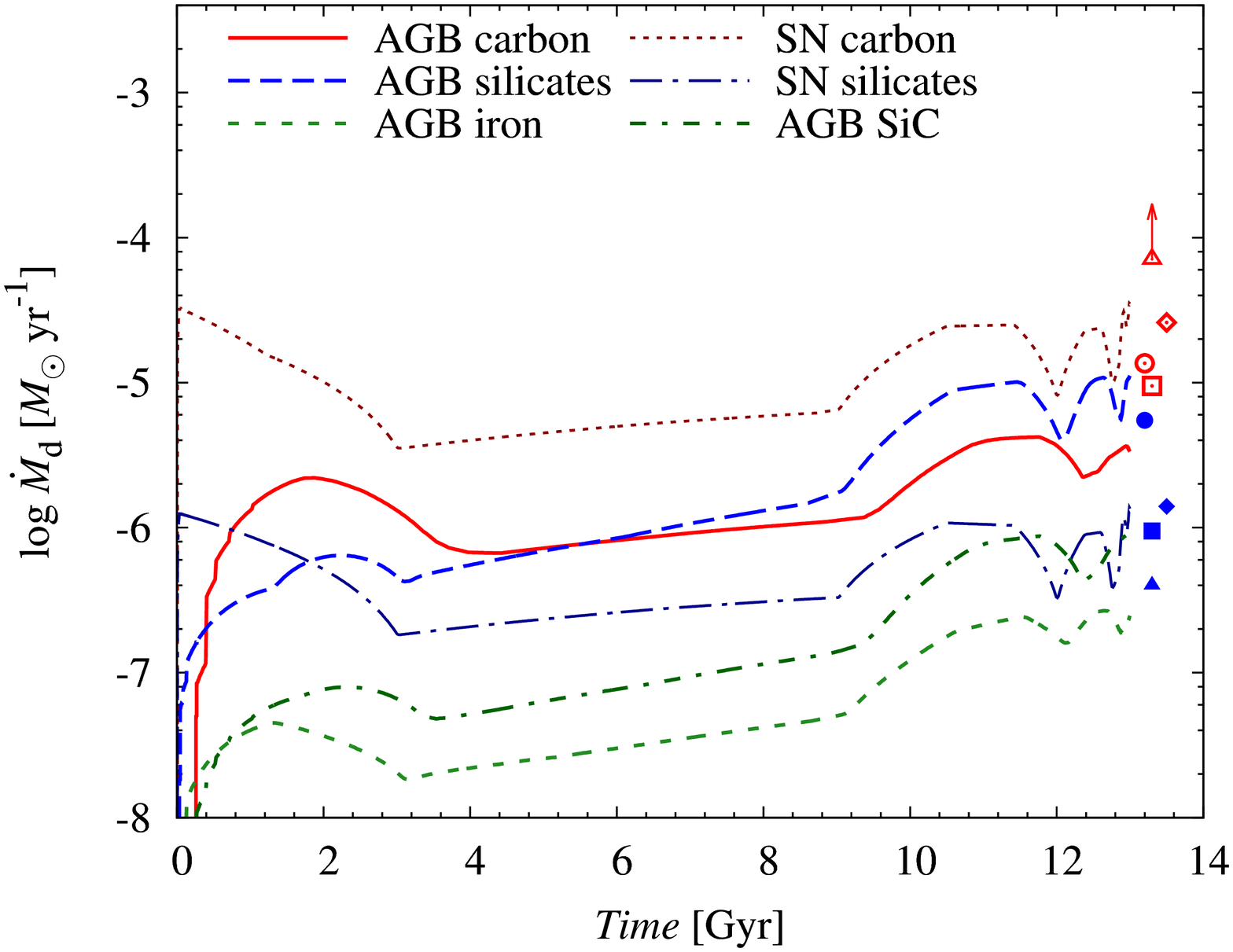}
	    \caption{The same as in Fig.~\ref{fig:DustRates}, but with dust yields from \citet{Ventura:2012cs, Ventura:2012gl}.}
	\label{fig:DustRates-vent}
\end{figure}

Thus, the total mass of dust of type $j$ per unit stellar mass returned to the ISM by evolved stars of mass $m$ at instant $t_d$, which survived until time $t$ is
\begin{equation}
\label{eq:Mdustsurv}
    M^{\rm s}_{j,\rm d}(m,t_{\rm d}, t) = f_{\rm AGB}(m,t_{\rm d})  Y_{j,\rm d}(m,Z)  {\rm e}^{\displaystyle -{(t-t_d)/\tau_{j,\rm dest}} }\,. 
\end{equation}
The total mass of stardust of type $j$ that has survived and accumulated by time $t$ is derived by integration over stellar masses and the whole evolution period
\begin{equation}
\label{eq:Mdusttot}
       M_{j,d}^{\rm tot} (t) = \int_{m_{\rm l}}^{m_{\rm WD}}dm' \int_0^t M^{\rm s}_{j,\rm d}(t,t',m')dt'\,,
\end{equation}
where the lower mass limit $m_{\rm l}$ is determined by the stellar lifetime $\tau=t$. The instant of stellar death $t_d$ can be easily converted to the initial stellar metallicity $Z$ by means of Eq.~(\ref{eq:Z_star}), which is more useful for comparison with observations. Therefore, in the following we will express $M^{\rm s}_{j,\rm d}(m,t_{\rm d}, t)$ as a function of $Z$ instead of $t_{\rm d}$.

The relative contribution of all stars with mass $m$ and initial metallicity $Z$ to the stardust population of type $j$ is described by the probability density
\begin{equation}
\label{eq:ProbSurv}
		P_{j,\rm d}(m,Z,t) = {M^{\rm s}_{j,\rm d}(t, Z, m)\over M_{j,d}^{\rm tot} (t)}  \,,
\end{equation}
which is the ratio of dust mass from these stars which has survived until instant $t$ to the total mass of this grain population.

\paragraph*{Input parameters.}
For numerical calculations of the probability density $P_{j,\rm d}(M,Z,t)$ for the dust species of interest we first calculate the mass distribution function from Eqs.~(\ref{eq:Z_star}) and (\ref{eq:FreqAGB}) using the best-fit star formation rate from \cite{Harris:2009p7602} and metallicity evolution from \cite{Carrera:2008p7783}, together with dust masses from Eqs.~(\ref{eq:Mdustsurv}) and (\ref{eq:Mdusttot}). The dust yields $Y_{j,\rm d}$ for AGB stars used throughout this work are taken from \cite{Ferrarotti:2006p993} and \cite{Zhukovska:2008bw}, unless it is stated otherwise. For the duration of dust-forming phase $\Delta t_{\rm AGB}$  in Eq.~(\ref{eq:FreqAGB}), we adopt the values from the model calculations of \cite{Ferrarotti:2006p993}, which start from the onset of TP-AGB evolution and stop when the envelope mass decreases below $0.01\Ms$. The duration of C-rich evolution for the number of carbon stars is calculated from the instant when the C/O ratio in the wind exceeds 1. Since the model calculations in \cite{Ferrarotti:2006p993} did not include stars with $m<1\Ms$, in this mass range we use the duration of TP-AGB phase from \cite{Girardi:2010ge} for $Z \simeq0.001$, and from \cite{Marigo:2007cw} for higher metallicities.

\section{Results}
\label{sect:Results}
\subsection{Evolution of dust and gas injection rates}
\label{sect:Results-DPRs}
The results of our calculations of the time evolution of the total injection rates for silicate, carbonaceous, SiC and metallic iron dust from AGB stars in the LMC are shown in Fig.~\ref{fig:DustRates}. Note that we assume that a circumstellar dust shell formed during the stellar mass-loss stage is ejected to the ISM instantaneously at the end of the AGB phase. 
For comparison, the figure also includes the DPRs from type II SNe calculated with condensation efficiencies of 0.15 for carbon, and 0.001 for silicate and iron grains \citep{Zhukovska:2008bw, Zhukovska:2008p7215}. 
AGB stars remain the major stellar site of dust formation over the entire LMC evolution, except for the first 200~Myr, during which carbon dust input from type II SNe dominates dust production. At the present time, AGB stars contribute $\sim70\%$ of the total stardust production rate, with the rest coming from type II SNe.  In this estimate we do not include dust formed at pre-SN stages in the circumstellar environment of red supergiants, Wolf-Rayet stars and luminous blue variables, because it is still unclear how much dust is condensed during these evolutionary stages, and what fraction survives subsequent SN explosion. It is likely that dust grains formed in outflows of a supernova progenitors are largely destroyed by thermal sputtering, when the forward SN shock from the explosion runs over the circumstellar shell \citep{Dwek:1992da}. 

The time evolution of the DPRs of AGB stars has two peaks, due to dust input from stellar generations formed during the initial and recent burst of star formation, separated by a much lower dust injection rate during the Age Gap. The peaks in the DPRs by AGB stars are shifted relative to the DPRs by SNe because of delayed dust injection from low- and intermediate-mass stars. 
 
Figure~\ref{fig:DustRates} demonstrates the variations of the dust mixture from AGB stars over time caused by the chemical evolution of the LMC. For the first several Gyrs, AGB stars return an almost pure carbon dust mixture  to the ISM. Later on, low- and intermediate-mass stars formed at higher metallicities enter the AGB stage of evolution and produce some silicate and iron grains, but their production rate increases significantly only with the onset of dust formation by stellar populations enriched by the recent star formation burst.
\begin{figure}[!tb]
\vspace*{-3cm}
\hspace*{-.7cm}
   	\includegraphics[scale=.35]{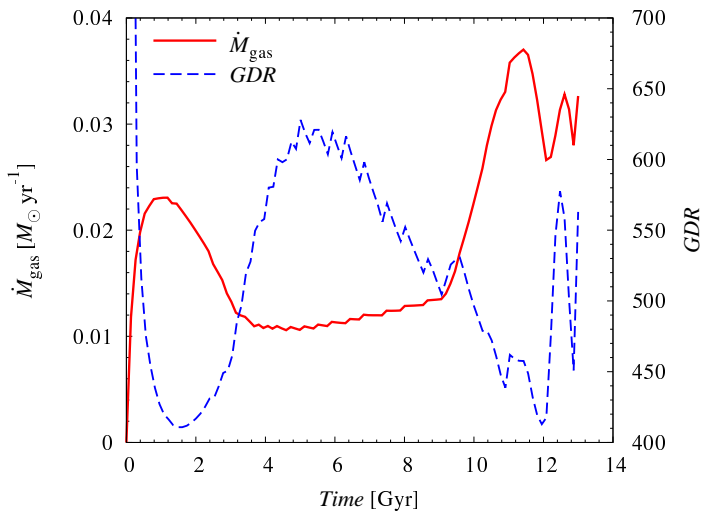}
	\caption{Variations of the total gas injection rate by AGB stars (solid line, vertical axis on the left) over the LMC evolution. The total gas-to-dust mass ratio $GDR$ in stellar ejecta is also plotted in the figure (dashed line, vertical axis on the right).}
	 \label{fig:GasInjRate}
\end{figure}

In order to check how the dust input from AGB stars depends on the stellar evolution prescriptions, we also calculate DPRs with the recent  dust yields from \citet{Ventura:2012cs, Ventura:2012gl}. The resulting DPR for carbonaceous, silicate, SiC and iron dust species, together with the same calculations for type II SN dust input are shown in Fig.~\ref{fig:DustRates-vent}. Dust production of silicates is much more efficient with these dust yields, so that it dominates the dust mixture after 5~Gyr, in contrast to findings from observations (see discussion in Sect.~\ref{sect:DPR_compar}). The injection rates of iron and SiC dust do not depend on metallicity as critically as in the work by \cite{Zhukovska:2008bw} and starts from $\sim10^{-7}\Msyr$. At the early epoch of LMC evolution silicate formation is inefficient because of low metallicities, while the carbon dust production rate is 10 times lower compared to the value calculated with dust yields from  \cite{Zhukovska:2008bw}. For this reason,  with the dust yields from \citet{Ventura:2012cs, Ventura:2012gl} the dust input from stars is dominated by type II SN over the entire LMC history, even with the relatively low condensation degrees in SNe assumed here. The total dust mass contributed by AGB stars is lower compared to the calculations with the dust yields from \cite{Zhukovska:2008bw}, particularly at early times. 

The time evolution of the gas injection rate from the AGB stellar population in the LMC is shown in Fig.~\ref{fig:GasInjRate}.  Here we also show the evolution of the average gas-to-dust ratio in the stellar ejecta of AGB stars derived by dividing the total injection rate of gas by that of dust. Stellar ejecta of AGB stars are the most dust-rich  with gas-to-dust ratios dropping down to 400 during the bursts of star formation and increases during the Age Gap. The reason for this is that low-mass stars with $m \lesssim 2\Ms$ formed during the early burst of star formation constitute a substantial fraction of the AGB stellar population during the Age Gap. These stars have lower dust-to-gas ratios in their ejecta compared to AGB stars in the mass range 2--4\Ms{} (Fig.~\ref{fig:agbdust2gas}). Note that the total gas-to-dust ratio from AGB population in the LMC is determined by ejecta from C-stars. O-rich AGB stars with metallicities $\lesssim Z_{\rm LMC}$ have significantly higher gas-to-dust ratio than C-stars. In total, AGB stars eject $2\times 10^8\Ms$ of gas to the ISM over 13~Gyr of evolution of the LMC. For comparison, the present interstellar gas mass is $7.9 \pm 0.7 \times 10^8\Ms$.

\begin{figure}[!tb]
		\vspace*{-3.5cm}
	\includegraphics[scale=.37]{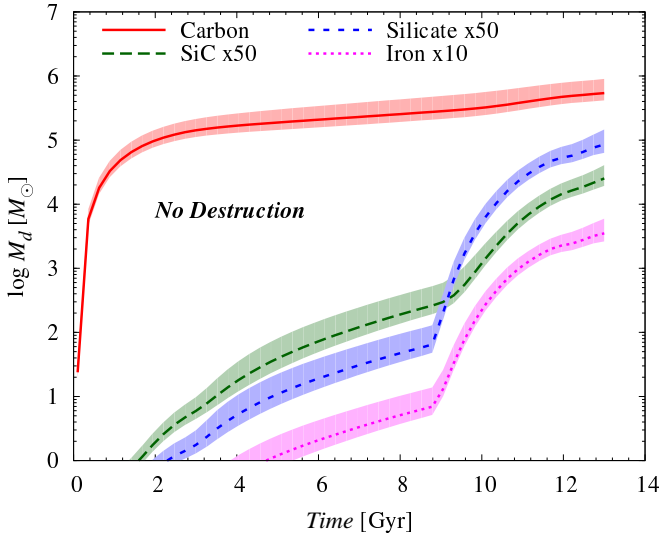} 
	\caption{Time evolution of accumulated dust mass ejected by AGB stars over the LMC history for carbonaceous, silicate, SiC and iron grains \textit{without} any destruction processes. Uncertainties in dust masses for each species due to uncertainties in the SFH are shown as shaded regions.}
\label{fig:DustMass}
\end{figure}

\begin{figure}[!tb]
		\vspace*{-3.5cm}
	\includegraphics[scale=.37]{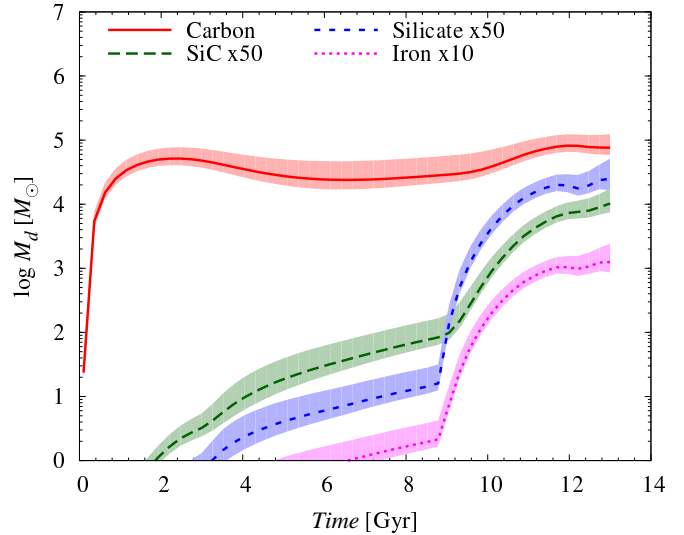}
	\caption{The same as Fig.~\ref{fig:DustMass}, but with grain destruction in the ISM on a time scale of 1.1~Gyr for carbon, SiC, and iron grains and 0.8~Gyr for silicates. }
\label{fig:DustMass-dest}
\end{figure}

\subsection{Accumulation of stardust mass}
The fraction of grains from AGB stars in the interstellar dust budget strongly depends on the efficiency of dust destruction in the ISM. We present the evolution of the total mass of stardust grains in the LMC without and with destruction in Figs.~\ref{fig:DustMass} and \ref{fig:DustMass-dest}. Figures also show uncertainties in accumulated stardust masses due to uncertainties in SFH.
The variations of total dust masses calculated without grain destruction in the ISM, i.e. cumulative dust masses, for various species formed by AGB stars over cosmic time provide an upper limit for the contribution from low- and intermediate-mass stars. In reality, the  contribution of AGB stars to the galactic dust budget is smaller than this value, because even in the absence of dust destruction in SN shocks, grains are inevitably destroyed in the ISM by the star formation process on a time scale shorter than the age of the LMC, as discussed in Sect.~\ref{sect:Destr}. 

The cumulative mass of carbon grains grows rapidly in the early burst of star formation, followed by a gradual increase to the present value of $5 \times 10^5 \Ms$ (Fig.~\ref{fig:DustMass}). Grain destruction on a time scale of $0.8-1.1$~Gyr keeps the carbon stardust mass at the level of several $10^4\Ms$ over the whole LMC evolution, with an increase up $10^5\Ms$ during the last Gyr. SiC is the second most abundant dust species from AGB stars until the end of the Age Gap 4~Gyr ago. For the last 4~Gyr silicates are the most abundant species after carbon; iron and SiC grains have comparable masses in the ISM, which are only $\lesssim 0.1$\% of the carbon dust mass. We performed additional calculations with the two times shorter grain lifetimes corresponding to the upper limit of the present-day SN II rate. The resulting value of accumulated dust mass from AGB stars is $4.5\times 10^4\Ms$.

\begin{figure}[!tb]    
		\vspace*{-3.cm}
	\includegraphics[scale=.37]{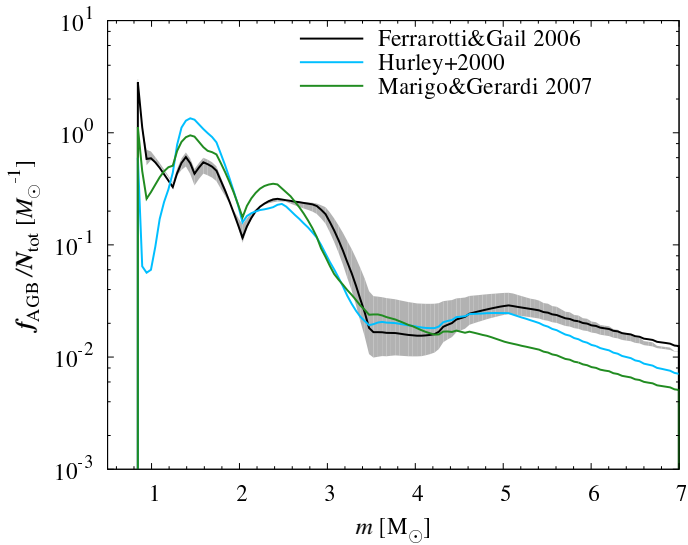}
	\caption{Stellar mass distribution of the present day population of TP-AGB stars relative to the total number of TP-AGB stars. Shaded area shows variations due to uncertainties in star formation history.  Results calculated for the duration of TP-AGB phase from \cite{Marigo:2007cw} and \cite{Hurley:2000gx} are shown for comparison (see explanation in the text).}
\label{fig:FreqAGB}
\end{figure}

\begin{figure}[!tb]    
		\vspace*{-3.cm}
	\includegraphics[scale=.37]{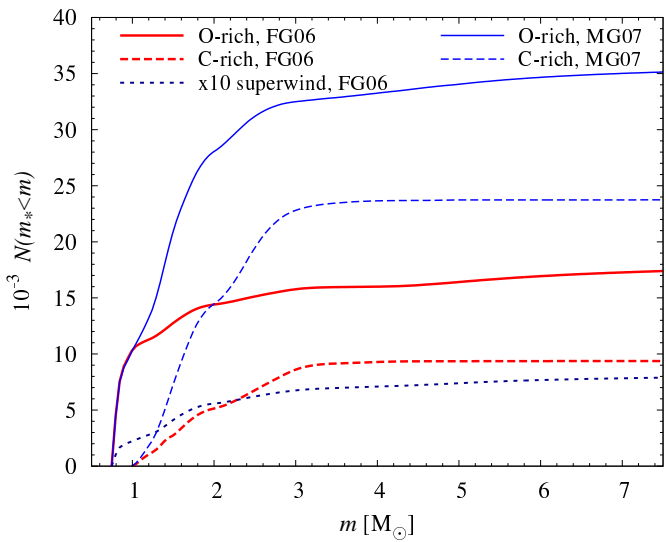}
	\caption{Cumulative stellar mass distribution showing the number of TP-AGB stars below mass $m$.
	}
\label{fig:NAGB}
\end{figure}

\subsection{Present population of dust-forming AGB stars}
We calculate the stellar mass distribution of the present population of TP-AGB stars from Eq.~(\ref{eq:FreqAGB}), taking the age of the LMC for the instant of stellar death $t_{\rm d}=13$~Gyr. The resulting stellar mass distribution of TP-AGB stars, normalised to the total number of these stars $N_{\rm AGB}^{\rm tot}$,  is shown in Fig.~\ref{fig:FreqAGB} as a function of the initial stellar mass. The mass distribution function has a peak at $\sim 0.8\Ms$ corresponding to the oldest and most numerous population of low-mass stars. Stars below $1~\Ms$ produce small amounts of dust and are neglected for the dust input here, but we account for the gas input from these stars in Sect.~\ref{sect:Results-DPRs} and consider their contribution to the mass distribution function. Another peak at $1.5\Ms$ ($Z=0.006$) corresponds to the dominant population of dust forming AGB stars with masses $1\lesssim m \lesssim 2\Ms$ and metallicities $0.003\lesssim Z \lesssim 0.007$. In order to illustrate the effect of uncertainties in the SFR, Fig.~\ref{fig:FreqAGB} also shows the stellar mass distribution calculated for the lower and upper limits of the SFH. The largest variations in $f_{\rm AGB}$ appear for the intermediate-mass stars as a result of larger uncertainties in star formation for the last 100~Myr. Because of this reason the relative fraction of AGB stars with $m>2.5\Ms$ vary from 18\% to 26\%.

The mass distribution and the total number of TP-AGB stars is very sensitive to the adopted durations  $\Delta t_{AGB}$ of this evolutionary stage. For comparison, Fig.~\ref{fig:FreqAGB} shows the results of calculations with $\Delta t_{AGB}$ from \cite{Ferrarotti:2006p993}, \cite{Marigo:2007cw}, and from analytical stellar evolution formulas \citep{Hurley:2000gx}. For $m<1\Ms$ and $Z=0.001$,  all model calculations assume TP-AGB durations from \cite{Girardi:2010ge}, which are based on the most recent stellar evolution calculations and have lower values compared to those from \cite{Marigo:2007cw}. The total number of TP-AGB stars $N_{\rm AGB}^{\rm tot}$ derived from Eq.~(\ref{eq:NAGB}) is 26\,870 for the best-fit SFR with variations of 19\,810 to 46\,070 for the lower and upper limits of star formation for $\Delta t_{AGB}$ from \cite{Ferrarotti:2006p993}. For the TP-AGB durations from \cite{Hurley:2000gx} and \cite{Marigo:2007cw}, the corresponding numbers of stars are 29\,000 and 54\,000 for the best-fit SFR.

The current numbers of oxygen- and carbon-rich TP-AGB stars in the LMC are shown in Fig.~\ref{fig:NAGB} as a cumulative function of the initial stellar mass.  For the duration of oxygen- and carbon-rich stellar evolution from the model calculations of \cite{Ferrarotti:2006p993}, we derive the total number of O-rich  stars to be 17\,500 for the best-fit SFR, and values of 13\,000 and  29\,170 for the lower and upper limits, respectively. The total number of carbon stars is about 9\,370 with a lower limit of 6\,810 and an upper limit of 16\,900. Because of the longer durations of TP-AGB  and  C-rich evolution derived by \cite{Marigo:2007cw}, the numbers of these stars are larger by about a factor of two. 

With respect to the evolutionary stages of the present day population of AGB stars, observational studies distinguish a small population of highly evolved AGB stars (extreme AGB stars), which comprise only several per cent of the sample by number, but account for 75\%-97\% of the total DPR \citep[][]{Riebel:2012eq, Boyer:2012ck}. For a rough estimate of the number of highly evolved stars from our model, we calculate the number of stars in the superwind phase adopting the value of $30 \times 10^3$~yr from  \citep{Schroder:1999ti} for the duration $\Delta t_{\rm AGB}$ of dust-forming phase in Eq.~(\ref{eq:FreqAGB}). We find that 800 stars out of the present TP-AGB stellar population of the LMC are in the superwind phase. This is two times lower than the number of extreme carbon-rich AGB stars found by \cite{Riebel:2012eq}, but very similar to the value of 886 stars found by \cite{Boyer:2012ck}.

\begin{figure}[tb]
	\vspace*{-3.cm}
	\includegraphics[scale=.37]{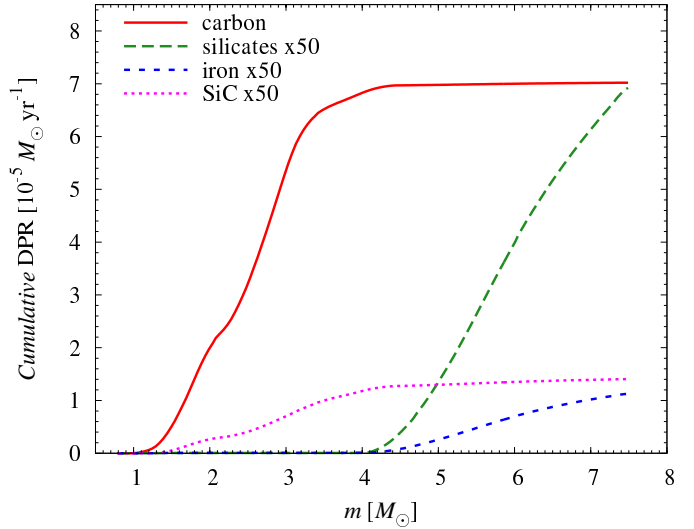}
	\caption{Cumulative dust production rate from AGB stars vs. initial stellar mass for present time.}
\label{fig:DPR_cum}
\end{figure}

\begin{figure}[tb]
\vspace*{-3 cm}
		\hspace*{-.5cm}
\includegraphics[scale=.38]{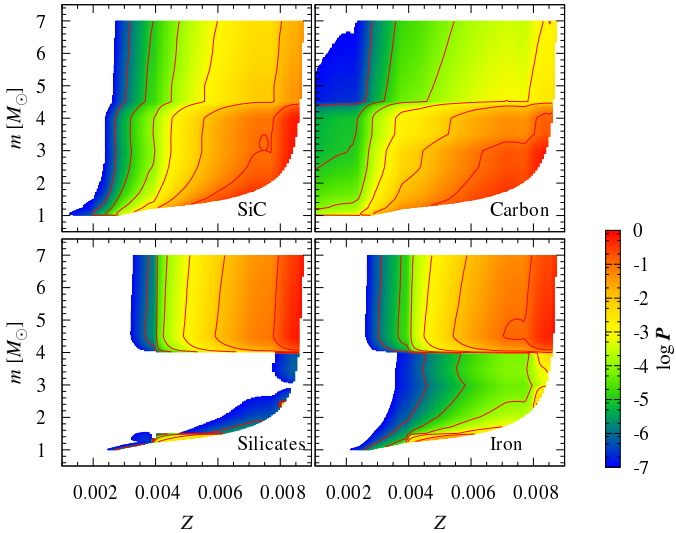}
	\caption{Probability density that an AGB star of initial stellar mass $M$ and metallicity $Z$ contributes to the current populations of stardust for various grains in the ISM of the LMC. The adopted time scales of dust destruction in the ISM are 0.8~Gyr for silicate grains and 1.1~Gyr for carbonaceous, SiC grains, and iron grains.}
\label{fig:Prob_LMC1Gyr}
\end{figure}

\begin{figure}[tb]
\vspace*{-2.5cm}
		\hspace*{-.5cm}
\includegraphics[scale=.38]{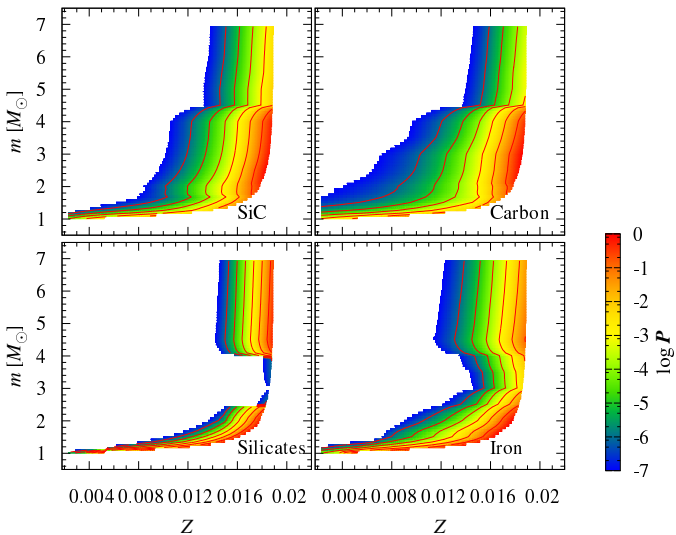}
	\caption{The same as Fig.~\ref{fig:Prob_LMC1Gyr}, but for the Solar neighbourhood. The time scales of dust destruction in the ISM are 0.4~Gyr for silicate grains and 0.6~Gyr for carbonaceous, SiC, and iron grains.}
\label{fig:Prob_MW}
\end{figure}


The contribution of the present population of TP-AGB stars of various masses to the DPR is illustrated in Fig.~\ref{fig:DPR_cum}, which shows the cumulative rate of dust production as a function of the initial stellar mass. This figure demonstrates that most carbonaceous and SiC dust grains originate from stars with $m<4\Ms$, while most silicate and metallic iron grains originate from stars with $m\gtrsim 4 \Ms$. In the mass range below $4\Ms$ only stars with mass slightly above $1\Ms$ contribute some iron and silicate dust at a rate, which is of the order of $\sim 1\%$ of the total DPR for these dust species. Recently \cite{McDonald:2011eo, McDonald:2010cy} suggested that metallic iron grains can be efficiently condensed in the stellar winds of AGB stars with masses below $1\Ms$ and low metallicities. Since the dust yields used here are calculated for stars with $m \geqslant 1\Ms $, an estimate of the contribution of stars with $m < 1\Ms $ to the total dust input requires new model calculations of dust condensation in stellar wind for stars in this mass range. However, we do not expect a high condensation degree of iron into metallic iron grains at low metallicities for the same reasons, which were discussed for inefficient SiC condensation at low metallicities in Sect.~\ref{Sect:DustModels}.

\subsection{Individual contributions of AGB stars to stardust populations}
In the following we quantify the relative contribution from all stars formed  over the LMC evolution to the present-day grain populations in terms of probability densities. The results of calculations of probability densities $P_{d,j}(m,Z)$  for a grid of the initial stellar masses and metallicities are shown in Fig.~\ref{fig:Prob_LMC1Gyr}. 
In order to illustrate the effect of metallicity on the parent stellar population of stardust grains, we also show $P_{d,j}(m,Z)$ for the Solar neighbourhood in Fig.~\ref{fig:Prob_MW}. The probability densities were calculated for the present time using the chemical evolution model of the Solar vicinity from \cite{Zhukovska:2008p7215}, where all physical quantities were averaged in a cylinder with the galactocentric radius of the Solar System.

The right and lower edges of the probability density distributions show the mass and metallicity of stars ending their life at the present time. It is almost vertical for stellar masses $m \gtrsim 2.5\Ms$, because the lifetimes of these stars are shorter than the time scales for galactic chemical enrichment of the LMC, and therefore the metallicity is close to the current metallicity of the LMC. In contrast, the current population of AGB stars with $m \lesssim 2.5\Ms$ demonstrate a large spread in initial metallicities. Stars that end their life close to the present time have the highest $P(m,Z)$, because their grains spent the least time in the ISM and have the highest chance to survive. The pace of grain destruction in the ISM also influences the span of the probability density distributions over $Z$, since it limits the time available for accumulation of stardust mass. For example, the range of initial metallicities of stars that contributed to the current stardust populations is much narrower for our galaxy than that in the LMC because of the shorter time scales of dust destruction (Figs.~\ref{fig:Prob_LMC1Gyr} and \ref{fig:Prob_MW}).

In the following we consider the individual stellar contributions to current populations of various grains.

\paragraph*{Carbon and Silicon Carbide grains.} 
Most of carbonaceous and SiC grains from AGB stars originate from stars in the mass range 1.5--4.4$\Ms$, in both the LMC and the Milky Way. The envelopes of these stars become C-rich after a number of thermal pulses, and enable efficient  carbon dust condensation in the stellar outflow, from C atoms synthesized in the He-burning shell. Low-mass stars with $m \lesssim 1.5\Ms$ are the most numerous, but they produce less C-rich dust than higher mass AGB stars \citep{Ferrarotti:2006p993}, and thus have a smaller total contribution. 

A comparison of $P(m,Z)$ for carbon and SiC grains reveal that stars with $Z\lesssim0.002$ have contributed only to the carbon dust population, since SiC condensation requires Si atoms (which are not synthesised in AGB stars) and can proceed efficiently only at higher metallicities.

\paragraph*{Silicate and iron grains.}
The probability densities for contributions from AGB stars to silicate and iron stardust populations in the LMC and Solar vicinity are shown in Fig.~\ref{fig:Prob_LMC1Gyr} and \ref{fig:Prob_MW}, respectively. The main  difference in their distributions is that silicate and iron grains in the LMC originate mainly from AGB stars with initial masses $\gtrsim 4$~\Ms, which constitute only 4\% of all AGB stars by number. The mass distribution of stars contributing to the silicate grain population in the Solar neighbourhood is bimodal, with approximately equal fractions coming from low- and high-mass AGB stars. The iron dust production is dominated by the most numerous population of AGB stars, with $1\Ms\lesssim m \lesssim2.5\Ms$. This discrepancy demonstrates the strong dependence of the dust composition from AGB stars on the initial metallicity resulting not only from a smaller amount of silicate-forming atoms (O, Si, Mg, Fe) available at low Z, but also from differences in the stellar evolution of low-mass stars at the AGB phase (see Sect.~\ref{sect:AGBdustyields}).  



The population of AGB stars with $m\gtrsim 4\Ms$ and $Z \simeq Z_{\rm LMC}$ dominate the silicate and iron dust production in the LMC, mainly because the dust yields for stars in this mass range with metallicities $Z \lesssim Z_{\rm LMC}$ by far exceed those from low-mass stars (Fig.~\ref{fig:dustyields}). Additionally, numerous low-mass stars in the LMC ending their life at present time have metallicities $\lesssim 1/2 Z_{\rm LMC}$, for which the returned masses for silicate and iron dust are even lower than for $Z \simeq Z_{\rm LMC}$.  The relative contribution of intermediate-mass stars to silicate and iron dust production  decreases at higher  metallicities, since low-mass stars experience a larger number of thermal pulses and spend a longer time as M-stars. The metallicity of the majority of AGB stars which contributed to the current stardust grain population in the Solar vicinity exceeds $Z_{\rm LMC}$. At these metallicities, both low- and intermediate-mass stars can produce substantial amounts of silicate and iron grains, which explains the bimodal mass distribution of the parent population of the silicate grains. Figures~\ref{fig:Prob_LMC1Gyr} and \ref{fig:Prob_MW} indicate that the relative fractions of iron grains from low-mass stars are larger than those of silicates, for both the LMC and Milky Way. This is also related to the condensation models from \cite{Ferrarotti:2006p993}, in which metallic iron can condense in stellar winds with C-rich chemistry. 


\section{Discussion}
\label{sect:Discussion}

\subsection{Current dust production rates: comparison with observations}
\label{sect:DPR_compar}



The Magellanic Clouds are unique objects that permit comparison between the observationally measured  DPR for the entire generation of evolved stars with the theoretically calculated dust input rates. 
However, such comparison has to be done with caution. In our calculations,  the total dust injection rate given by Eq.~(\ref{eq:DustInjRate}) is determined by a number of stars ending their life per unit time, which instantaneously eject the entire dust mass condensed during the AGB evolution into the ISM. Whereas observations probe the mass-loss rates of AGB stars at different evolutionary stages, and may underestimate the total dust input rates to the ISM. Stars experience moderate dust mass loss for most of the AGB evolution, with the exception of brief increases during luminosity spikes following thermal pulses. A period of very efficient dust formation, which occurs  during  the superwind phase at the tip of AGB, is relatively short, typically about 30\,000~yr \citep[e.g.,][]{Schroder:1999ti}. 

 The current dust injection from AGB stars derived from our calculations is by far dominated by carbon-rich dust (97\% of the total DPR), with only a minor fraction of oxygen-rich dust ($\lesssim 2\%$, Table~\ref{tab:LMC_DPR}). This result confirms findings from observational studies of dust input from evolved stars in the LMC \citep[][]{Zijlstra:2006gl, Matsuura:2009p363, Riebel:2012eq, Boyer:2012ck}. The predicted production rate for carbon dust of $5.7\times 10^{-5}\Msyr$ agrees well with an estimate derived by \cite{Matsuura:2009p363} and is about two times the value obtained by \cite{Srinivasan:6p7509}. It is somewhat higher than the most recent estimates derived with the GRAMS radiative transfer model fitting by \cite{Riebel:2012eq} and from $8\mum$ excess derived by \cite{Boyer:2012ck} (about 4 and 6 times higher, respectively). 

There is a large scatter between the observationally-derived values of the DPR for oxygen-rich dust (Table~\ref{tab:LMC_DPR}). Our calculations predict the production rate of O-rich dust of $1.3 \times 10^{-6}$~\Msyr, which agrees well with the value from \cite{Srinivasan:6p7509}. It is 1.5 times higher than the value derived by \cite{Boyer:2012ck} and 4 times lower than the DPR measured by \cite{Riebel:2012eq}. Our calculations probably underestimate dust production by O-rich AGB stars due to inefficient dust condensation at low-mass rates in the models adopted here (see Sect.~\ref{Sect:DustModels}), therefore our value provides a lower limit for DPR from O-rich stars.

Some differences in the global DPR from various studies are related to variations in the total number of dust-forming stars (Table~\ref{tab:LMC_DPR}). Our calculated number of carbon stars is 5.2 times greater than the value measured by \cite{Matsuura:2009p363} and only 1.3--1.5\footnote{For the number of C-rich sources from work of \cite{Srinivasan:6p7509}, we combined the number of regular C-rich stars and extreme AGB stars assuming that the latter population is dominated by C-rich stars.} times greater than the values measured in other studies in Table~\ref{tab:LMC_DPR}. 
The differences in the number of carbon stars are not sufficiently large to explain the differences of a factor 4--6 in carbon dust injection rate. Possible reasons for this will be discussed in the next paragraphs. The variations in the number of oxygen-rich stars from different studies are greater compared to the number of carbon-rich stars. Our calculated number is agrees within 10\%  with the number of O-rich stars in the sample of \cite{Riebel:2012eq} and \cite{Boyer:2012ck}, and 2 times greater than the value derived by \cite{Srinivasan:6p7509}.

It is important to remember that photometric techniques do not constrain the chemical compounds of the dust mixture. Therefore, the DPRs for all stars are derived with some representative fixed dust properties. Variations in optical constants for different dust compositions are one of the most important sources of uncertainties and can cause differences in DPR of a factor of 4--7 \citep{Boyer:2012ck}, which is similar to the differences between our calculated DPR and those measured observationally.

Another potential source of differences between theoretical and observational estimates of the DPR originate from different assumptions for the dust shell expansion velocities. The observational studies summarised in Table~\ref{tab:LMC_DPR} assume a fixed value of the expansion velocity of  $10\kms$ for all AGB stars. In \cite{Ferrarotti:2006p993} the expansion velocity is calculated from the equation of momentum conservation for dust-driven winds, and varies from a few $\kms$, to much higher values of several $10\kms$ at the final thermal pulses. In the case of low mass-loss rates \cite{Ferrarotti:2006p993} assume  a constant value of the expansion velocity of $5\kms$. Since the total dust input is dominated by highly evolved AGB stars, the lower value of expansion velocity assumed in observational studies, compared to \cite{Ferrarotti:2006p993}, may lead to lower values of the DPR. However, the resulting difference should not exceed a factor of two. 
 
Metallic iron grains are not included in the analysis of observational studies of the global dust injection rates, as it has no easily identifiable spectroscopic features, and  is expected to be a minor component. We find that it indeed contributes only $0.4\%$ by mass to the grain population from AGB stars. 

Calculations with dust yields from \citet{Ventura:2012cs, Ventura:2012gl} predict a dust mixture dominated by silicates, in contrast to what is observed in the circumstellar shells of AGB stars in the LMC (Fig.~\ref{fig:DustRates-vent}). The current DPR of silicates is 2.5 time higher than that of carbon dust, and several times higher than the value derived by \cite{Riebel:2012eq}. The discrepancy of the calculated input rate of O-rich dust compared to other estimates from observations is about an order of magnitude. The current DPR for carbon dust is significantly lower than all estimates from observational works. A possible explanation for this discrepancy may be that the efficiency of the hot bottom burning, which prevents a star from becoming a carbon star, is overestimated by \citet{Ventura:2012cs, Ventura:2012gl}. \cite{Marigo:2007dc} found that the hot bottom burning may be weakened, or even prevented, in the intermediate-mass stars with $m\approx 3-4 \Ms$ and metallicities $Z \gtrsim 0.001$, due to variable molecular opacities in the stellar envelope when C/O approaches unity. Since the mass fraction of stars in this mass range is about one third of all AGB stars producing  O-rich dust, this effect might significantly increase the average carbon-to-silicate ratio in the dust mixture from AGB stellar populations.


\subsection{Stardust vs. interstellar dust}
With models employing star formation history, we can compare the grain population from AGB stars accumulated over the LMC evolution with existing interstellar dust. The dust mass in the ISM can be estimated using the average dust-to-gas mass ratio and total gas mass from observations. 
Dust emission studies found that the average gas-to-dust ratio in the LMC is 2-4 times the value in the Solar neighbourhood
 \citep[][]{Bernard:2008gv, RomanDuval:2010ic, 2011A&A...536A..17P, Galliano:2011gy}. This is very similar to the values of the gas-to-dust ratio that have been inferred from extinction studies \citep[e.g.,][]{Weingartner:2001p475, Welty:2012ct, Gordon:2003p7510}. For the value of the Galactic gas-to-dust ratio of 170  \citep{Zubko:2004p4116}, we derive an average gas-to-dust ratio in the LMC of 340 - 680, and a total dust mass of $1.1-2.5 \times 10^6\Ms$.  For this estimate, we adopted a total gas mass in the LMC of $7.9 \pm 0.7 \times 10^8\Ms$ \citep{Israel:1997tm,StaveleySmith:2003bl}.

Our calculations of the grain life-cycle with dust destruction yield a current mass of dust from AGB stars of $8.3\times10^4\Ms$  (Fig.~\ref{fig:DustMass}), which is about 13 - 30 times lower than the existing ISM dust mass. It is somewhat larger than an estimate of accumulated stardust mass of $5\times 10^4\Ms$, based on dust injection rates from IR observations  \citep{Matsuura:2009p363}.  The reason for this is that \cite{Matsuura:2009p363}  adopted a shorter dust lifetime $\tau_d$ of 400~Myr typical for the Milky Way. If we adopt the same value for $\tau_d$, corresponding to the upper limit of SN II rate in the LMC, we derive a lower value of the accumulated stardust mass from AGB stars of $4.5\times 10^4\Ms$. In addition to dust destruction in SN blast waves and in star formation considered here, grains can be removed from the ISM by galactic winds. There is diverse evidence that the LMC may be loosing its gas via galactic winds or energetic outflows powered by star formation feedback \citep{Carrera:2008p7783,StaveleySmith:2003bl,Lehner:2007be, Nidever:2008cz}, although this issue is still debated. 

The total stardust mass from the calculations without any dust destruction in the ISM ($5.3\times 10^5\Ms$, Fig.~\ref{fig:DustMass}) is still about 2-4.3 times smaller than the current interstellar dust mass. Thus, even without including any destruction processes, there is a discrepancy between the dust input from AGB stars and existing dust mass in the ISM of the LMC. 
Additional dust growth by collection of dust-forming atoms in the ISM on existing interstellar grains, which is very important in the Milky Way, is also required to reproduce the observed dust mass in the LMC (Zhukovska et al, in preparation).

\cite{Zhukovska:2008bw} found that, in our galaxy, the role of stars in dust production increases at lower metallicities. For an accumulated dust mass of $8.3\times10^4\Ms$ and an adopted value of the gas mass of the LMC, the mass fraction of dust from AGB stars in the LMC is about $3.4-7.7\%$, which is indeed higher than the corresponding value of about $1.5-2\%$ in the Solar neighbourhood  \citep{Zhukovska:2008bw}. With a metallicity of about $0.2\, \Zs$, the Small Magellanic Cloud offers an opportunity to further investigate the dependence of the stardust fraction on metallicity, which will be considered in a future paper.  


\section{Conclusions}
\label{sec:Concl}

We investigated the dust input from AGB stars in the LMC over cosmic time scales with a model of dust evolution and mass- and metallicity-dependent dust yields. The most recent constraints on the star formation history and metallicity are used to construct the populations of AGB stars and their evolution in the LMC.

AGB stars remain the major stellar source of dust production over the entire LMC history, except the first 200~Myr, during which type II SNe  are the dominant dust source. At the metallicity of the LMC and lower, the stardust population is dominated by carbon grains formed mostly out of primary carbon. Carbon dust production by AGB stars has two broad peaks, 12 and 2 Gyr ago, with dust return rates of $7-10\times 10^{-5}\Msyr$ separated by a period of a lower dust production rate of $\sim 2\times 10^{-5}\Msyr$ during the Age Gap. In contrast, the formation of silicates is very inefficient, until the end of the Age Gap. It rapidly increases to its present level of $\sim 10^{-6}\Msyr$ with dust return from the generation of AGB stars formed out of the ISM enriched by recent bursts of star formation. Iron and SiC dust production shows, similar to silicates, time variations at much lower rate.
For the present epoch, we compare theoretically calculated dust production rates by AGB stars to the values derived from IR observations, and find a generally good agreement within the scatter of various observational estimates.  Our predicted carbon dust production rate of $6.7\times 10^{-5}\Msyr$ is closer to the highest values from observations. The silicate dust production rate of $1.3\times 10^{-6}\Msyr$ agrees well with the values from studies using IR color excess, but is several times lower than estimates made with pre-computed radiative transfer models.


We quantitatively characterise the contribution of AGB stars to the populations of the main stardust species as a function of initial stellar mass and metallicity. The majority of carbon and SiC grains originates from the most numerous group of stars with masses $\lesssim 4\Ms$ and metallicity from slightly below $Z_{\rm LMC}$ to  $\approx0.5 Z_{\rm LMC}$. The parent stellar population of present-day silicate and iron stardust grains in the LMC consists of intermediate mass stars with $m \gtrsim 4\Ms$ and metallicity close to $Z_{\rm LMC}$, which constitute only $\lesssim 4$\%  of stars by number, in contrast to the bimodal mass distribution of these stars in the Solar neighbourhood. This indicates the strong dependence of the dust formation in AGB stars on the initial metallicity, resulting not only from a smaller amount of dust-forming atoms, but also from different stellar evolution of low-mass stars at the AGB stage at low metallicity. Because of differences in the stellar mass distribution (and, consequently, ages) the present population of dusty oxygen-rich stars in the LMC is more metal rich than carbon-rich stars.

The cumulative dust mass produced by AGB stars over cosmic time scales from our model calculations is $5\times 10^{5}\Ms$, which is two times lower than the existing dust mass in the LMC. Taking into account grain destruction in the ISM the mass of dust that survived until present is only $8.3\times 10^{4}\Ms$. Thus, even without including any destruction processes in the ISM, there is a discrepancy between dust input from AGB stars and existing dust mass in the ISM of the LMC. We conclude that dust growth by mantle accretion in the ISM, a major dust source in the Solar neighbourhood, is also important at subsolar metallicities of the LMC.

\section*{Acknowledgements}
SZ acknowledges support by the \textit{Deutsche Forschungsgemeinschaft} through SPP 1573: ``Physics of the Interstellar Medium''. We thank the anonymous referee for constructive comments and suggestions that improved the paper. We thank Hans-Peter Gail and Paul Boley for their careful reading of the manuscript. SZ thanks Margaret Meixner and STScI for their hospitality during an extended visit, when this work was begun. 


\end{document}